\documentclass[aps,prd,amsmath,showpacs,nofootinbib]{revtex4}

\usepackage{epsfig}
\input psfig.sty

\usepackage{graphicx}% Include figure files
\usepackage{dcolumn}% Align table columns on decimal point
\usepackage{bm}% bold math
\usepackage[english]{babel}
\usepackage{color}

\newcommand{\mytilde}{\raise.17ex\hbox{$\scriptstyle\mathtt{\sim}$}}

\newcommand{\dd}{{\rm d}}

\newcommand{\aapr}{Astron. Astrophys. Review}
\newcommand{\mnras}{MNRAS}
\newcommand{\aap}{Astron. and Astrophys.}

\newcommand{\no}{\mbox{$n_{e_o}$}}

\title[Galaxy clusters, type Ia supernovae and the fine structure constant]

\begin{document}
\title{Galaxy clusters, type Ia supernovae and the fine structure constant}

\author{R. F. L. Holanda$^{1,2,3}$\footnote{E-mail: holanda@uepb.edu.br}}

\author{ V. C. Busti$^4$\footnote{E-mail: viniciusbusti@gmail.com}}

\author{L. R. Cola\c{c}o$^2$\footnote{E-mail: colacolrc@gmail.com}}

\author{J. S. Alcaniz$^5$\footnote{E-mail: alcaniz@on.br}}

\author{S. J. Landau$^6$\footnote{E-mail: slandau@df.uba.ar }}

\address{$^1$Departamento de F\'{\i}sica, Universidade Estadual da Para\'{\i}ba, 58429-500, Campina Grande - PB, Brasil}

\address{$^2$Departamento de F\'{\i}sica, Universidade Federal de Campina Grande, 58429-900, Campina Grande - PB, Brasil}

\address{$^3$Departamento de F\'{\i}sica, Universidade Federal do Rio Grande do Norte, 59300-000, Natal - RN, Brasil}

\address{$^4$Departamento de F\'{\i}sica Matem\'{a}tica, Instituto de F\'{\i}sica, Universidade de S\~{a}o Paulo, 
CP 66318, CEP 05508-090, S\~{a}o Paulo - SP, Brazil}

\address{$^5$Observat\'orio Nacional, 20921-400, Rio de Janeiro - RJ, Brasil}

\address{$^6$Departamento de F\'{\i}sica, Facultad de Ciencias Exactas y Naturales, Universidad de Buenos Aires and IFIBA, CONICET, Ciudad Universitaria - PabI, Buenos Aires 1428, Argentina}

\begin{abstract}

As is well known, measurements of the Sunyaev-Zeldovich effect can be combined with observations of the  X-ray surface brightness of galaxy clusters to estimate the angular diameter distance to these structures. In this paper, we show that this technique depends on the fine structure constant, $\alpha$. Therefore, if $\alpha$ is a time-dependent quantity, e.g., $\alpha=\alpha_0 \phi(z)$, where $\phi$ is a function of redshift, we argue that current data do not provide the real angular diameter distance, $D_{\rm{A}}(z)$, to the cluster, but instead $D_A^{data}(z) = \phi(z)^2 D_{\rm{A}}(z)$. We use this result to derive constraints on a possible variation of $\alpha$ for a class of dilaton runaway models considering a sample of 25 measurements of $D_A^{data}(z)$ in redshift range $0.023 < z < 0.784$ and estimates of $D_{\rm{A}}(z)$ from current type Ia supernovae observations. We find no significant indication of variation of $\alpha$ with the present data.

\end{abstract}

%\pacs{} %{98.80.-k, 98.80.Es, 98.65.Cw}

\maketitle
\section{Introduction}

Galaxy clusters (GC) are the most massive bound systems  in the Universe, containing hundreds to thousands of galaxies, and several cosmological information can be extracted from their observations. From the evolution of GC X-ray temperatures and their X-ray luminosity function one may put limits on the matter density parameter, $\Omega_m$, and the normalisation of the density fluctuation power spectrum, $\sigma_8$  (Henry 2000; Ikebe {\it{et al.}} 2002; Mantz {\it{et al.}} 2008; Vanderline {\it{et al.}} 2010). The  dark energy equation-of-state parameter, $w$, can  be constrained by the abundance of galaxy clusters as a function of mass and redshift , with statistical errors competitive with other techniques (Albrecht {\it{et al.}} 2006, Basilakos, Plionis \& Lima 2010; Chandrachani Devi \& Alcaniz, 2014). The gas mass fraction via X-ray observations of galaxy clusters can also be used to constrain cosmological parameters (Sasaki  1996; Pen 1997; Ettori {\it{et al.}}  2003; Allen {\it{et al.}} 2008; Lima {\it{et al.}} 2003; Gon\c calves 
{\it{et al.}} 2012). 

An important phenomena occurring in galaxy clusters is the Sunyaev-Zel’dovich effect (SZE) (Sunyaev \& Zeldovich, 1972),  a small distortion of the cosmic microwave background spectrum (CMB), due to the inverse Compton scattering of the CMB photons passing through a population of hot electrons. Such effect  is independent of the galaxy cluster redshift, being a powerful tool to study the Universe at high-$z$. In particular, combined observations of the SZE and X-ray emission of the intracluster medium (Sarazin, 1988) have been used to extract important information in cosmological and fundamental physics. 

Some years ago, Uzan { {{\it{et al.}}}} (2004) argued that the SZE/X-ray technique can be used to measure angular diameter distances $D_A$ to galaxy clusters and test the so-called cosmic distance duality relation (CDDR) (Etherington, 1933; Ellis, 1971)
\begin{equation}
{D_L \over D_A(1+z)^{2}}=\eta=1\;, 
\end{equation}
a widely used relation in observational cosmology ($D_L$ is the luminosity distance), which is valid when source and observer are connected by null geodesics in a Riemannian spacetime and the number of photons is conserved. More recently, Holanda, Gon\c{c}alves \& Alcaniz (2012) showed that measurements of the gas mass fraction via SZE and X-ray  of galaxy clusters  can also be used to test the hypotheses behind the above relation. By allowing deviation of the CDDR such as  $D_L/[D_A(1+z)^{2}]=\eta(z)$,  where $\eta$ is an arbitrary function of $z$, no compelling evidence for violation was found (see Table I of Holanda, Busti \& Alcaniz (2016) for a summary of recent estimates of $\eta(z)$). 

The time and/or spatial variation of fundamental constants is a prediction from theories that attempt to unify the four fundamental interactions. In the last decade the issue of the variation of fundamental constants has experienced a renewed interest, and several astronomical observations and local experiments have been performed to study  their possible variations (Uzan 2011, GarciaBerro 2007) and to establish limits on such variations. Local experiments include geophysical methods such as the natural nuclear reactor that operated about $1.8\times 10^9$ years ago in Oklo, Gabon (Damour \& Dyson 1996; Petrov et al. 2006, Gould et al 2006; Onegin 2012) and laboratory measurements  of atomic clocks with different atomic numbers (Fisher et al. 2004, Peik et al. 2004, Rosenband et al 2008).  On the other hand,  torsion balance experiments (Wagner et al 2012) and the Lunar Laser Ranging experiment (Muller et al 2012) can also give 
indirect bounds on the present spatial variation of $\alpha$ (Kraiselburd \& Vucetich 2012) or on the present time variation of $\alpha$.

The astronomical methods are based mainly on the analysis of high-redshift quasar absorption systems. The many-multiplet method,  which compares the characteristics of different transitions in the same absorption cloud, is the most successful method employed so far to measure possible variations of $\alpha$ (Webb 1999). Most of the reported results are consistent with a null variation of fundamental constants. Nevertheless, Murphy et al.  (Murphy et al. 2003, King et al. 2012) have claimed a detection of  a possible spatial variation of $\alpha$ using Keck/HIRES and VLT/UVES observations. However a recent analysis  of the instrumental systematic errors (Whitmore \& Murphy 2015) of the VLT/UVES data rules out the evidence for a space or time variation in $\alpha$ from quasar data. On the other hand, constraints on the variation of $\alpha$ in the early universe can be obtained by using the available 
Cosmic Microwave Background (CMB) data (O'Bryan et al 2015, Planck collaboration 2015) and the abundances of the light elements generated during the Big Bang Nucleosynthesis (Mosquera \& Civitarese 2013). A possible  time variation of the fine structure constant $\alpha$ was also tested from combined observations of the SZE and X-ray emission in GC.  For example, Galli (2013) proposed a method using  the integrated Comptonization parameter $Y_{SZ}D^2_A$ and its X-ray counterpart, the $Y_X$ parameter. The ratio of the two parameters was shown to be dependent on the fine structure constant as $\alpha^{3.5}$. More recently, Holanda { {{\it{et al.}}}} (2016) argued that measurements of the gas mass fraction can also be used to probe a time evolution of $\alpha$.  In this latter reference, it was shown that observations of the gas mass fraction via the SZE and X-ray surface brightness of the same galaxy cluster are related by 
\begin{equation}
f_{SZE} = \phi(z)\eta(z)f_{X−ray}\;,
\end{equation}
where $\phi(z) =\alpha/\alpha_0$ (the subscript ``0" denotes the value of a quantity at the present time). Taking into account a direct relation, shown by Hees {\it{et al.}} (2014), between variation of $\alpha$ and the CDDR, i.e.,   $\phi(z)=\eta^2(z)$, and considering a class of dilaton runaway models in which $\varphi(z)=1-\gamma \ln{(1+z)}$, it was found  $\gamma = 0.037 \pm 0.18$, consistent with $\gamma=0$ at $1\sigma$ level. 

In this paper, we deepen our analysis of the relation between $\alpha$ and $\eta$ and show that the measurements of $D_A$ to galaxy clusters from SZE/X-ray observations also depend  on  the fine structure constant and, therefore, on a possible time-dependence of this quantity. As in our previous work, we assume the theoretical framework of the runaway dilaton proposed by Damour et al. (Damour {\it{et al.}} 2002a; 2002b;) to model variations in $\alpha$. We show that if $\alpha=\alpha_0 \phi(z)$, which necessarily implies $\eta(z) \neq 1$, current SZE and X-ray observations do not provide the real angular diameter distance but instead $D^{data}_A(z)=\phi(z)\eta^2(z)D_A(z)$.  We use current type Ia supernovae (SNe Ia) data to obtain the true distances to galaxy clusters. Then we combine SNe Ia data  with 25 $D^{data}_A(z)$ measurements from De Filippis { {{\it{et al.}}}} (2005) in the redshift interval $0.023 < z_{GC} < 0.784$ to place bounds on the time-dependence of $\phi(z)$ for a  class of dilaton runaway models (Damour {\it{et al.}} 2002a; 2002b; Martins {\it{et al.}} 2015). 

The paper is organised  as follows. In Sec. II we discuss in detail the method proposed. The data samples used in our analysis 
are discussed in Sec. III. In Sec. IV we perform our analyses and discuss the results. We end the paper with our main conclusions in Sec. V.

\section{Method}

\subsection{$D_A$ from SZE and X-ray observations}

As it is well known, SZE and X-ray surface brightness can be used  to obtain the angular diameter distance to galaxy clusters. The SZ effect can be quantified as  
\begin{equation}
 \Delta T_{\rm SZ}(\theta) =f(\nu, T_e)
 \frac{kT_e T_0}{m_ec^2}\sigma_T\int_{-\ell_{\rm max}}^{\ell_{\rm max}}
 n_e \dd\ell
\end{equation}
where we assume an isothermal model. In the above expression, $n_e$ is the electronic density of the intracluster medium, $T_e$ is the electronic temperature,
$k_B$ the Boltzmann constant, $T_0$ = 2.728K is the present-day temperature
of the CMB, $m_e$ the electron mass and $f(\nu, T_e)$ accounts for frequency
shift and relativistic corrections (Itoh, Kohyama \& Nozawa
1998; Nozawa, Itoh \& Kohyama 1998). $\ell_{\rm max}$  (actually, $2\ell_{\rm max}$) is
the length of the path along the line of sight inside the halo of
the cluster whereas  $\theta$ is  the angular distance from
the clusters centre projected on the celestial sphere. The Thompson cross section is written in terms of the fine structure constant ($\alpha=e^2/c \hbar$) as
\begin{equation}
\sigma_T=\frac{8\pi \hbar^2 }{3 m_e^2 c^2}\alpha^2.
\end{equation}
where $e$ is the electronic charge, $\hbar$ is the reduced Planck constant %divided by $2\pi$, 
and $c$ is the speed of light.

 The X-ray emission, on the other hand, is due to thermal
bremsstrahlung and the surface brightness is given by
\begin{equation}
 S_X(\theta) =
 \frac{1}{4\pi}\frac{D_A^2}{D_L^2}\int_{-\ell_{\rm
 max}}^{\ell_{\rm max}} \frac{\dd L_x}{\dd V}\dd\ell,
\end{equation}
where the emissivity in the frequency band $[\nu_1,\nu_2]$  can be written in terms of the fine structure constant as
\begin{eqnarray}
 \frac{dL_x}{dV}=\alpha^3\left(\frac{2\pi k_BT_e}{3m_e}\right)^{\frac{1}{2}}\frac{2^4\hbar^2}{3 m_e}n_e\left(\sum_i{Z_i^2n_ig_{Bi}}\right)\;.
 \label{eq3.17}
\end{eqnarray}
In the above expression,  $Z_i$ and $n_i$ are, respectively, the atomic numbers and the distribution of elements and $g_B$ is the Gaunt factor which takes into account the corrections due quantum and relativistic  effects of Bremsstrahlung emission. By considering the intracluster medium constituted  by  hydrogen and helium we can write
\begin{equation}
 \frac{\dd L_X}{\dd V} = \Lambda_e n_e^2,
\end{equation}
where $\Lambda_e$ is the so-called X-ray cooling function, which is proportional to $\alpha^3$.

Now, let us assume the $\beta$-model for the galaxy cluster, where the electron density of the hot
intra-cluster gas has a profile of the form (Cavaliere \& Fusco-Fermiano 1978)
\begin{equation}\label{eqprof}
 n_e(r) =
 n_0\left[1+\left(\frac{r}{r_c}\right)^2\right]^{-3\beta/2},
\end{equation}
for $0<r<R_{\rm cluster}$ and 0 otherwise ($R_{\rm cluster}$ being
the maximum extension of the cluster). Introducing the angle $\theta_c$
\begin{equation}\label{thetac}
 \theta_c=r_c/D_A\;,
\end{equation}
where $r_c$ is the cluster core radius.

{  By using the profile given by Eq. (8), one can find  in the limit $R_{\rm
cluster}\rightarrow\infty$,

\begin{equation}
\label{eq:sz2} \Delta T_{\rm SZ} = \Delta T_0 \left( 1+
\frac{\theta^2 }{\theta_{c}^2} \right)^{1/2-3\beta/2},
\end{equation}
where $\Delta T_0$ is the central temperature decrement. More precisely:
\begin{equation}
\label{eqsze3} \Delta T_0 \equiv T_0
f(\nu, T_{\rm e}) \frac{ \sigma_{\rm T} k_{\rm B} T_{\rm e}}{m_{\rm e}
c^2}n_{e0} \sqrt{\pi} \theta_c D_A g\left(\beta/2\right),
\end{equation}
with
\begin{equation}
g(\alpha)\equiv\frac{\Gamma \left[3\alpha-1/2\right]}{\Gamma \left[3
\alpha\right]}, \label{galfa}
\end{equation}
where $\Gamma(\alpha)$ is the gamma function.

For X-ray surface
brightness, we have
\begin{equation}
S_X = S_{X0} \left( 1+ \frac{\theta }{\theta_{c}^2} \right)^{1/2-3
\beta}, \label{eqsxb1}
\end{equation}
where the central surface brightness $S_{X0}$ reads
\begin{equation}
\label{eqsxb2} S_{X0} \equiv \frac{D_A^2 \Lambda_{e}}{D_L^2 4 \sqrt{\pi} } n_{e0}^2 \theta_c D_A \ g(\beta).
\end{equation}

Therefore, assuming a possible time variation of $\alpha$, equations (3), (4), (5) and (6) clearly show that % if $\alpha=\alpha_0\varphi(z)$ 
both the SZE and X-ray data are directly affected by a possible departure from $\alpha_0$. 
\begin{figure*}
\centering
\includegraphics[width=3.5in, height=2.9in]%[width=0.49\textwidth]
{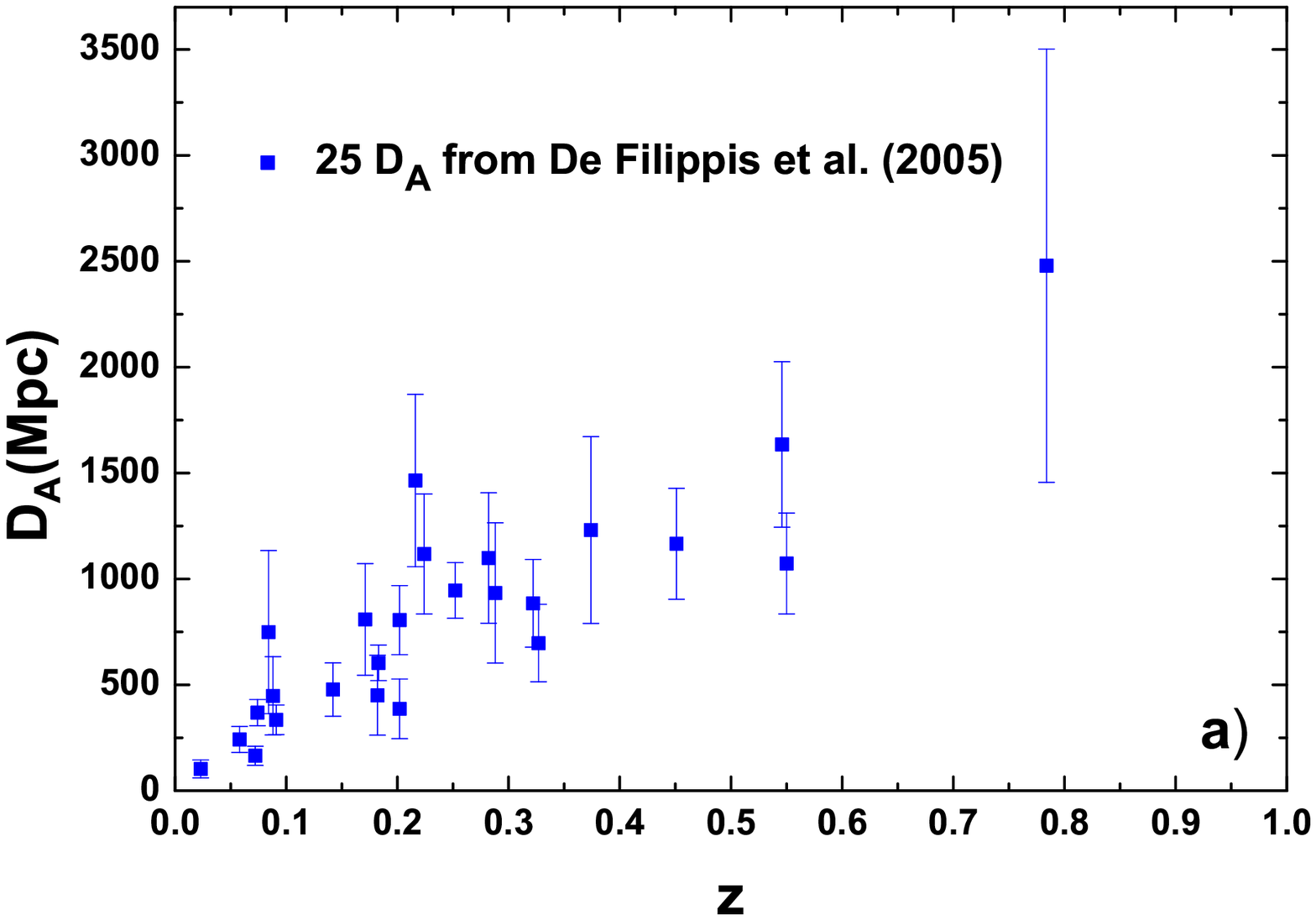}
%\hspace{0.3cm}
\includegraphics[width=3.5in, height=2.9in]{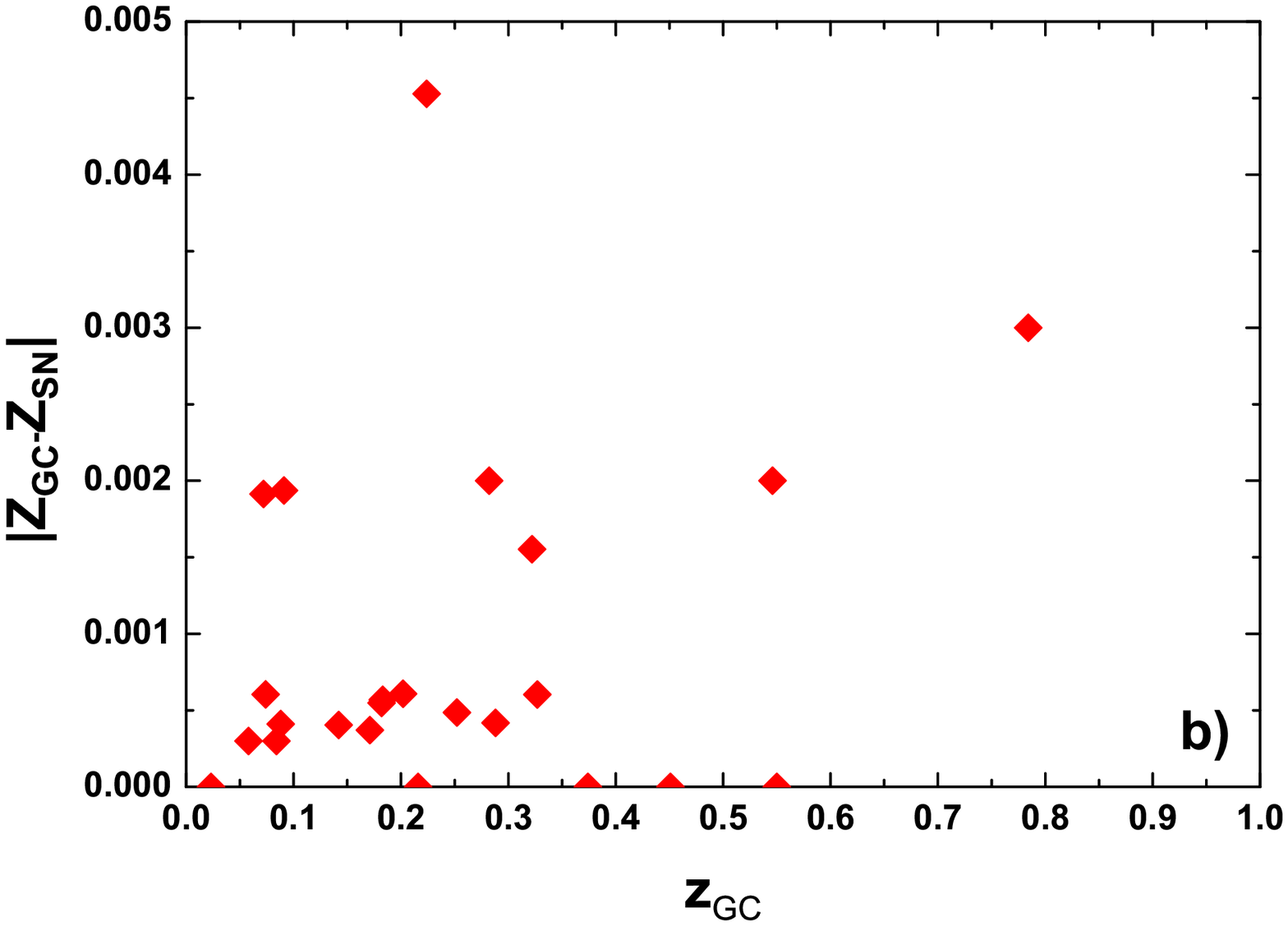}
\caption{(a) The blue square points are the observed angular diameter distances from De Filippis {\it{et al.}} (2005). (b) In this figure we plot  $|z_{GC}-z_{SN}|$ for the 25 pairs of galaxy clusters and SNe Ia used in our analysis.}
\end{figure*}

So, one can solve equations (11) and (14) for the angular diameter
distance by eliminating $n_{0}$, taking for granted the validity
of the CDDR and the constancy of $\alpha$. However, a more general result appears when these assumptions are relaxed. Considering  $\alpha=\alpha_0\phi(z)$ and $\eta \neq 1$, one obtains
\begin{widetext}
\begin{eqnarray}
\label{eqobl7}
{{D}}_A(z) &= & \left[ \frac{\Delta {T_0}^2}{S_{\rm X0}}
\left( \frac{m_{\rm e} c^2}{k_{\rm B} T_{e} } \right)^2
\frac{g\left(\beta\right)}{g(\beta/2)^2\ \theta_{\rm c}}
\right] \times %\nonumber \\
%& & \times
 \left[ \frac{\Lambda_e \phi^3(z)}{4 \pi^{3/2}f(\nu,T_{\rm
e})^2\ {(T_0)}^2 {\sigma_{\rm T}}^2\ (1+z_{\rm
c})^4}\frac{1}{\phi^4(z)\eta(z)^2} \right] \nonumber \\
\end{eqnarray}
\end{widetext}
The experimental quantity is given by

\begin{widetext}
\begin{eqnarray}
\label{eqobl7}
D_A(z)^{\: data} &= & \left[ \frac{\Delta {T_0}^2}{S_{\rm X0}}
\left( \frac{m_{\rm e} c^2}{k_{\rm B} T_{e} } \right)^2
\frac{g\left(\beta\right)}{g(\beta/2)^2\ \theta_{\rm c}}
\right] \times %\nonumber \\
%& & \times
 \left[ \frac{\Lambda_e }{4 \pi^{3/2}f(\nu,T_{\rm
e})^2\ {(T_0)}^2 {\sigma_{\rm T}}^2\ (1+z_{\rm
c})^4}\right].\nonumber \\
\end{eqnarray}
\end{widetext}
Therefore, instead of the real angular diameter distance, the currently measured quantity is $D_A^{\: data}(z)=D_A(z) \; \eta^2(z) \phi(z)$.}

\subsection{Relation between $\eta$ and $\alpha$}

Recently, Hees {{{\it{et al.}}}} (2014) investigated cosmological signatures of modifications of gravity via
the presence of a scalar field with a multiplicative coupling to the electromagnetic Lagrangian.  This
kind of coupling arises in various alternative theories of gravity (string theories, Kaluza-klein theories, among others). In a such framework, it was shown that variations of the fine structure constant and violations of the distance duality are intimately and unequivocally linked by 
\begin{equation}
\label{eq:dalpha_eta} 
\frac{\Delta \alpha(z)}{\alpha}=\frac{\alpha(z)-\alpha_0}{\alpha_0}=\eta^2(z)-1\;,
\end{equation}
which means that Eq. (\ref{eqobl7}) can now be written as %becomes
\begin{equation}
D_A(z)=D_A^{data}(z)\phi(z)^{-2}\;.
\label{eqphi}
\end{equation}

Most theories in which the local coupling constants become effectively  spacetime dependent,  involve some  kind of fundamental field (usually a scalar field) controlling such dependence. As mentioned earlier, in the present paper we focus on the dilaton runaway models (see more details in Damour, Piazza \&  Veneziano 2002 and Martins { {{\it{et al.}}}}  2014). In this model,  the
runaway of the dilaton towards strong coupling prevents violations of the Equivalence Principle. Furthermore, the relevant parameter for studying the variation of $\alpha$  is the coupling of the dilaton field to hadronic matter. We are interested in the evolution of the dilaton at low redsfhits, $0.023 < z < 0.784$, and thus it is a reasonable approximation to linearize the field evolution, such as 
\begin{equation}\label{evolslow}
\frac{\Delta\alpha}{\alpha}(z)\approx\, -\frac{1}{40}\beta_{had,0} {\phi_0'}\ln{(1+z)}\;,
\end{equation}
where $\phi_0'= \frac{\partial \phi}{\partial \ln a}$ is the scalar field at present time and $\beta_{had,0}$ is the current value of the coupling between the dilaton and hadronic matter. This last equation is the one we will use to compare the model predictions with galaxy cluster data through the method discussed  below (for more details on the derivation of the above expression, we refer the reader to Holanda { {{\it{et al.}}}} (2016)).

\subsection{$D_A$ from SNe Ia data}

{  In our analysis, the angular diameter distance $D_A$ for each galaxy cluster is  obtained using SNe Ia  data of the Union 2.1 compilation (Suzuki {\it{et al.}} 2011) and the CDDR\footnote{{  It has been pointed out by Kraiselburd {\it{et al.}} (2015) that the peak luminosity of SNe Ia could be affected by  a variation in $\alpha$. These authors also show that the differences
in the peak bolometric magnitudes  of the Union 2.1 compilation due to possible variations in $\alpha$ are too small  compared with the leading terms intervening in the calculation of the luminosity distances of Type Ia supernova. Here, we do not consider this kind of effect in our analysis.}}. On the Hees et al. (2014) framework, the CDDR must  be modified accordingly, such as ${D_L (1+z)^{-2}/D_A} = \phi^{1/2}$ (see Eqs. (1) and (17)). Consequently, the angular diameter distance $D_A$ for each galaxy cluster is given by: ${D_A} = D_L (1+z)^{-2}/\phi^{1/2} $. }

 Now, using Eq. (\ref{eqphi}) we obtain
\begin{equation}
\phi(z)^{3/2}= \frac{D^{data}_A(z) (1+z)^2}{D_L(z)},
\label{eqphi2}
\end{equation}
{  where the  luminosity distance $D_L$ is obtained from the observed distance modulus $\mu$ through $D_L(z)=10^{(\mu(z)-25)/5}$ and  $\sigma^2_{D_L}=(\frac{\partial D_L}{\partial \mu} )^2 \sigma^2_{\mu}$. Note that the distance modulus was calibrated using light curve parameters and a fiducial wCDM model through the SALT II fitter (Guy et al. 2007). 
As the Union2.1 consists of several subsamples, Suzuki et al. (2012) allowed a different absolute magnitude value for each subsample thereby making the impact of the cosmological model negligible.}

\begin{figure*}
\centering
\includegraphics[width=3.5in, height=2.9in]%[width=0.5\textwidth]
{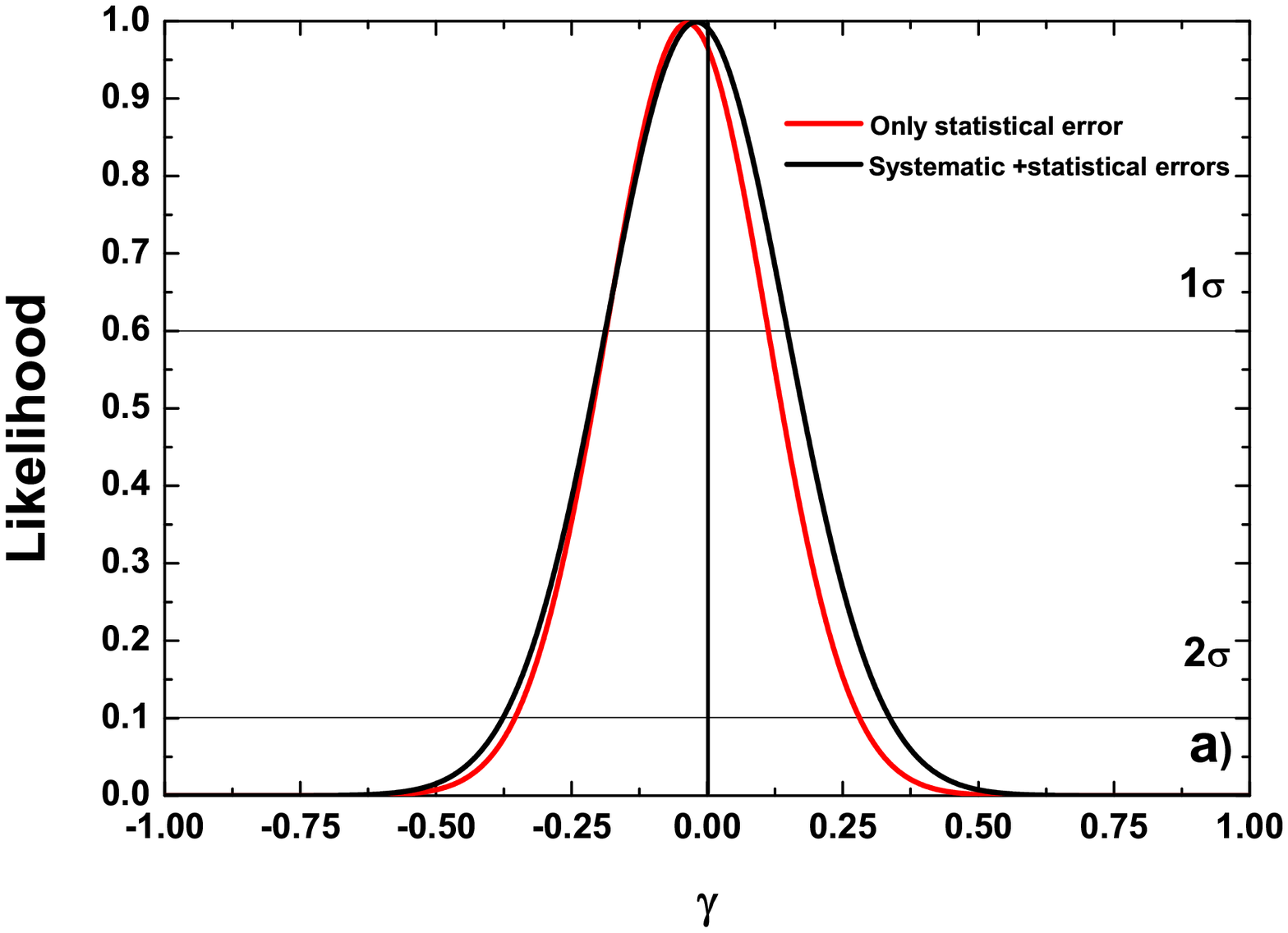}
%\hspace{0.3cm}
\includegraphics[width=3.5in, height=2.9in]%[width=0.5\textwidth]
{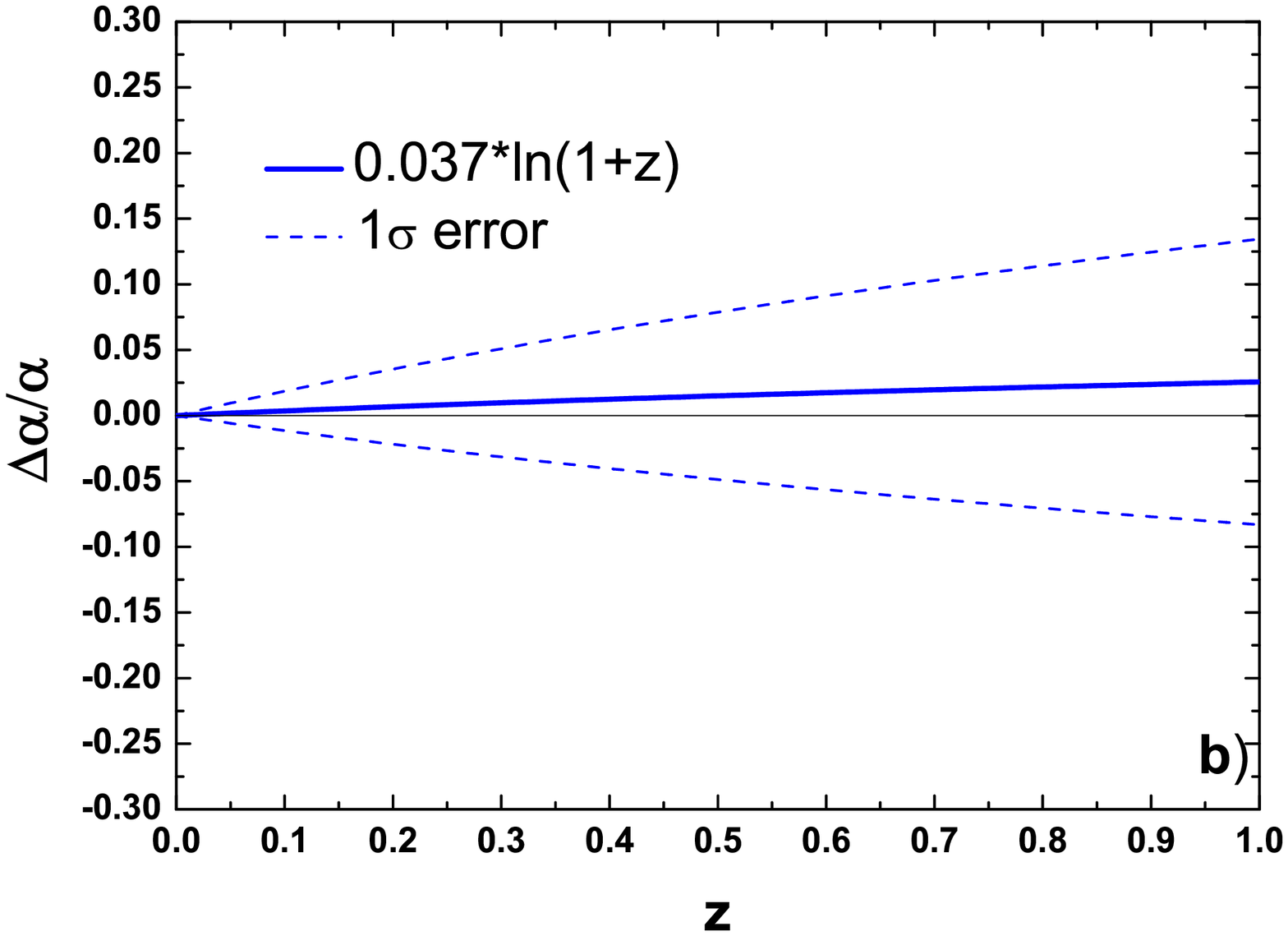}
\caption{ a) Constraints on a possible variation of the fine structure
constant. The red and black  solid  lines correspond to analysis from  angular diameter distances + SNe Ia with and without systematic error in galaxy cluster data, respectively. b) The evolution of $\Delta \alpha/\alpha$ from the best-fit values of our analysis.
}
\end{figure*}

\section{Galaxy cluster and SNe Ia Samples}

Motivated by the X-ray observations of Chandra and XMM Newton telescopes, which showed that in general galaxy clusters exhibit elliptical surface brightness maps,  De Filippis {\it{et al.}}  (2005) reanalysed, using an isothermal elliptical model,  two galaxy clusters samples for which combined X-ray and SZE analyzes have already been reported using an spherical model (Reese {\it{et al.}} 2002 and Ebeling {\it{et al.}} 1996). As a result, De Filippis {\it{et al.}} (2005) compiled measurements of the angular diameter distances to 25 galaxy clusters (see Fig. 1a) in redshift range $0.023 < z_{GC} < 0.784$. The choice of circular rather than elliptical model does not affect the resulting central surface brightness or Sunyaev-Zeldovich decrement and the slope $\beta$ differs slightly between these models. However,  different values for the core radius can be obtained, which change significantly the estimates of the angular diameter distance.

In order to perform our analysis, we also consider 25 measurements of the luminosity distance from the original 580 data points of Suzuki (2012), the so-called Union2.1 compilation. The redshifts of SNe Ia were carefully chosen to match the ones of galaxy clusters, with the larger redshift difference being smaller than 0.005 (see Fig. 1b). Before discussing our results, it is worth mentioning that  SZE observations are sensitive to  processes of energy injection into the cosmic microwave background radiation (CMB), that may change the standard evolution of the CMB temperature to $T_{CMB}(z) = T_{CMB}(z = 0)(1 + z)^{1+\beta}$. In principle, this effect should also be considered in SZE/X-ray technique. However, Saro { {{\it{et al.}}}} (2014) using SZE measurements of 158 galaxy clusters (at 95 and 150 GHz) from the South Pole Telescope (SPT) constrained the $\beta$ parameter to $\beta = 0.017 \pm 0.028$ ($1\sigma$), which is consistent with the standard model prediction ($\beta = 0$). {  Moreover, the frequency used to obtain the SZE signal in galaxy clusters sample of De Fillipis et al. (2005) was 30 GHz, in this band the effect on the SZE from a variation of $T_{CMB}$ is completely negligible. The best frequency is 150 GHz for negative signals and around 260 GHz for positive signals (Melchiorri \&  Melchiorri 2005).} Therefore, we do not consider a modified CMB temperature evolution law in our analysis.

\section{Analysis}

We evaluate our statistical analysis by defining the likelihood distribution function ${\cal{L}} \propto e^{-\chi^{2}/2}$, where
\begin{equation}
\label{chi2} 
\chi^{2} = \sum_{i = 1}^{N}\frac{{\left[(1-\frac{1}{40}\beta_{had,0} {\phi_0'}\ln{(1+z)}) - \phi_{i, obs} \right] }^{2}}{\sigma^{2}_{i, obs}},
\end{equation}
{$\phi_{i, obs} = (D^{data}_A (1+z)^2/D_L)^{2/3}$ and  {  the uncertainty associated to this quantity is  $\sigma^2_{\phi}=(\frac{\partial \phi}{\partial D_A} )^2 \sigma^2_{D_A} + (\frac{\partial \phi}{\partial D_L} )^2 \sigma^2_{D_L}$.} 
The sources of uncertainty in the measurement of $D_{A}^{\: data}(z)$ are: i) statistical contributions: SZE point sources $\pm 8$\%, X-ray background $\pm 2$\%, Galactic N$_{H}$ $\leq \pm 1\%$, $\pm 15$\% for cluster asphericity, $\pm 8$\% kinetic SZ and for CMB 
anisotropy $\leq \pm 2\%$, ii) estimates of systematic effects: SZ calibration $\pm 8$\%, X-ray flux calibration $\pm 5$\%, 
radio halos $+3$\% and X-ray temperature calibration $\pm 7.5$\%  (see table 3 in Bonamente { {{\it{et al.}}}} 2006). 
 
Constraints on the quantity $\gamma = \frac{1}{40}\beta_{had,0} {\phi_0'}$ are shown in Fig. (2a).  From this analysis we obtain $\gamma =-0.037\pm 0.157$ at 68.3\% (C.L.), which is full agreement with $\phi(z)=1$ or, equivalently, with no variation of fine structure constant $\alpha$. When we add the systematic errors to the  galaxy clusters data ($\approx 12\%$) we obtain $\gamma = -0.02\pm 0.17$ at 68.3\% (C.L.). These constraints on $\gamma $ can be compared with the limits obtained recently by some of us (Holanda { {{\it{et al.}}}} 2015). In this previous analysis it was shown that measurements of the gas mass fraction obtained via the Sunyaev-Zeldovich effect and X-ray surface brightness of the same galaxy cluster are related by $f_{SZE}=\phi(z)f_{X-ray}$, where $\phi(z)=\frac{\alpha}{\alpha_0}$. For 29 $f_{gas}$ measurements, reported by LaRoque { {{\it{et al.}}}} (2006), it was found $\gamma = 0.065\pm 0.095$ at 68.3\% (C.L.), which is in complete accordance with the results of the present analysis. 

 In Fig. (2b) we show our results for $\Delta \alpha/ \alpha =-\gamma \ln(1+z)$. Clearly, no significant evolution is verified. 
{  On the other hand, let us compare this result with other constraints reported in the literature with local and astronomical methods. The first ones constrain the present variation of $\alpha$ and can be divided into indirect and direct meaurements: i) Experiments designed to test the Weak Equivalence Principle (WEP) such as torsion balance tests and lunar laser ranging  and ii) Experiments designed to test the constancy of fundamental constants today such as the comparison of the frequencies of atomic clocks of different atomic number. The first ones constrain  the E\"{o}tv\"{o}s parameter $\zeta$\footnote{This parameter quantifies possible departures from exact equality between gravitational and inertial mass.} which in turn  can be related to the parameters of the dilaton model as follows: $\zeta \sim 5.2 \times 10^{-5} \beta_{had,0}^2$ (Martins {\it{et al.}} 2015) which for the current bounds on the WEP violation implies  $\beta_{had,0} \leq 10^{-4}$ (Murphy, Webb \& Flambaum 2003; Wagner et al. 2012). The most stringent bound on the  present time variation of the fine structure constant was obtained by Rosenband {\it{et al.}} 2008: $\frac{1}{\alpha} \frac{d\alpha}{dt} =(-1.6 \pm 2.3) \times 10^{-17} {\rm yr}^{-1}$. Assuming  the Hubble constant $H_0=67.4 \pm 1.4 {\rm km/s/Mpc}$, {  the constraint on the dilaton model parameters yields: $|\gamma| \leq 3 \times 10^{-5} $}. On the other hand, the astronomical method uses high-redshift quasar absorption spectra to put bounds on $\alpha$ variation  for a wide range of redshifts ( $0.1 < z < 3$). Martins {\it{et al.}} 2015 consider the data of King {\it{et al.}} 2012 and more recent data sets of dedicated measurements to put bounds on the dilaton model. Assuming the standard model value for the cosmological parameters they obtain the following bounds: $\beta_{had,0} \leq 10^{-4}$ and $\phi_0' \leq 10^{-2}$. 
In such way, the bounds obtained in this paper are weaker than bounds obtained with other methods. It is worth mentioning, however, that the method proposed in this work relies on a combination of physics and phenomena at different cosmic  (intermediate redshifts) and uses the well established physics of galaxy clusters.
 }
%Note also that the limits derived in this work are less stringent than others derived by local and astronomical methods, such as, torsion balance tests and lunar laser ranging, that provide $\beta_{had,0} \leq 10^{-4}$ (Muller {\it{et al.}} 2012; Wagner {\it{et al.}} 2012) and  atomic clocks with different atomic number, whose current bound is $\gamma \leq 3 \times 10^{-5}$ (Rosenband {\it{et al.}} 2008).(Martins {\it{et al.}} 2015). 

\section{Conclusions}

Measurements of the Sunyaev-Zeldovich effect can be combined with observations of the  X-ray surface brightness of galaxy clusters to estimate the angular diameter distance to these structures and impose limits in several cosmological parameters (see, e.g., Birkinshaw, 1999). In this paper we have shown that this technique depends on the fine structure  constant, $\alpha$, and, therefore, if $\alpha$ is a time-dependent quantity, e.g., $\alpha=\alpha_0\phi(z)$, current data do not provide the true distance $D_A(z)$, but instead $D_A^{data}(z)=\phi(z)^2D_A(z)$. 

In order to perform our analysis we have transformed 25 measurements of $D_L$ from current SNe Ia observations into $D_A(z)$, taking into account the direct relation, shown by Hees { {{\it{et al.}}}} (2014), between a variation of $\alpha$ and the CDDR. When combined with 25 measurements of $D_A^{data}(z)$  from galaxy clusters in the redshift range $0.023 < z < 0.784$, these data sets imposed cosmological limits on $\phi(z)$ for a class of dilaton runaway models. We have found $\Delta \alpha/ \alpha = (0.037 \pm 0.157)\ln(1+z)$, which is consistent with no variation of $\alpha$.

Finally, it is important to emphasize that although the limits obtained here on $\Delta \alpha/ \alpha$ are less stringent than others derived from different methods (e.g., quasar spectroscopy, atomic clocks and CMB) they rely on completely different physical mechanisms and redshift range (intermediate redshifts).  Moreover, the SZE/X-ray technique is independent of any calibrator usually adopted in the determinations of the distance scale. Since several SZE surveys are underway, the method discussed in this paper may be useful in the near future and reinforces the interest in the search for a possible time variation of the fine structure constant (and all its physical consequences) using the combination of measurements of the Sunyaev-Zeldovich effect and X-ray surface brightness of galaxy clusters.

%----------------------------------------------------------%
\section*{Acknowledgments}
%----------------------------------------------------------%

RFLH acknowledges financial support from INCT-A and CNPq (No. 478524/2013-7, 303734/2014-0). VCB  is  supported  by
Sao Paulo Research Foundation (FAPESP)/CAPES agreement  under  grant  number  2014/21098-1. SL is supported  by PIP 11220120100504  CONICET. L. R. Cola\c{c}o is supported by CAPES.


\begin{thebibliography}{99}
\bibitem{ad}Ade, P. A. R., et. al., 2015, Planck collaboration,
arXiv:1502.01589
\bibitem{al}Albrecht A., Bernstein G., Cahn R.,  {\it{et al.}} W. L. F., Dark Energy Task Force, ArXiv: 0609591, 2006
\bibitem{alen2}Allen S. W., Rapetti D. A., Schmidt R. W., Ebeling H., Morris R. G., Fabian A. C.,
 MNRAS, 2008, 83, 879
\bibitem{ba} Basilakos, S., Plionis, M., Lima, J. A. S., PRD, 2010, 82,  083517
\bibitem{birk}M. Birkinshaw, 1999, Phys. Rep., 310, 97,
\bibitem{bon}Bonamente M., Joy M. K., LaRoque S. J., Carlstrom J. E., Reese E. D., Dawson K. S., ApJ, 2006, 647, 25
\bibitem{cav}Cavaliere, A., \& Fusco-Fermiano, R. 1978, A\&A., 667, 70
\bibitem{Devi:2014rva} Chandrachani Devi, N. and Alcaniz, J. S., arXiv:1402.2590 [astro-ph.CO]
%\bibitem{chen} Chen Y. \&  Ratra, B., A\&A, 2012, 543, A104
%\bibitem{Marassi} Cunha, J. V., Marassi, L. \% J. A. S. Lima, MNRAS Lett., 2007, 379:L1
\bibitem{damour} Damour, T., \& Dyson, F.\ 1996, Nuclear Physics B, 480, 37 
\bibitem{d}Damour, T., Piazza, F. \& Veneziano, G., 2002a, PRD, 66, 046007 
\bibitem{d1} Damour, T., Piazza, F. \& Veneziano, G., 2002b, Physical Review Letters 89, 081601.
\bibitem{de}De Filippis, E., Sereno, M., Bautz, M.W., \& Longo, G. 2005, ApJ,
625, 108
\bibitem{et}Ettori S., Tozzi P., Rosati P., A\&A, 2003,  398, 879
\bibitem{et}Etherington, I. M. H. 1933, Phil. Mag., 15, 761; reprinted
in 2007, Gen. Relativ. Gravit., 39, 1055
\bibitem{fischer} Fischer, M., Kolachevsky, N., Zimmermann, M., et al.\ 2004, Physical Review Letters, 92, 230802 
\bibitem{galli}Galli, S., 2013, PRD, 87, 12, 123516
\bibitem{garciaberro} Garc{\'{\i}}a-Berro, E., Isern, J., \& Kubyshin, Y.~A.\ 2007, \aapr, 14, 113 
\bibitem{goncalves}Gon\c{c}alves, R. S., Holanda, R. F. L., Alcaniz, J. S., MNRAS, 2012, 420, L23
\bibitem{gould} Gould, C.~R., Sharapov, E.~I., \& Lamoreaux, S.~K.\ 2006, \prc, 74, 024607
\bibitem{guy}Guy, J., et al. 2007 A \& A, 466, 11 
\bibitem{hees} Hees, O., Minazzoli, A. \& Larena, J., 2014, PRD 90, 124064
\bibitem{holandacar} Holanda, R. F. L., Carvalho, J. C., Alcaniz, J. S., JCAP, 2013, 04, 027
%\bibitem{busti}  Holanda, R. F. L.,  Busti, V. C. \&  Pordeus da Silva, G., 2014, 443, L74
\bibitem{busti2}  Holanda, R. F. L.,  Busti, V. C. \&  Alcaniz, J. S., JCAP {1602}, no. 02, 054 (2016). [arXiv:1512.02486 [astro-ph.CO]].
\bibitem{holandaetal} Holanda, R. F. L.,  Landau, S. J.,  Alcaniz, J. S. ,  Sanchez G.,  I. E. \&  Busti, V. C. 2016, JCAP, 5, 047 
\bibitem{hin}Hinshaw, G. {\it{et al.}}, 2013, APJS, 208, 19H
\bibitem{it}Itoh, N., Kohyama, Y., \& Nozawa, S., 1998, ApJ, 502, 7-15
\bibitem{Ik}Ikebe Y., Reiprich T. H., Bohringer H., Tanaka Y., Kitayama T., A\&A, 2002,  383, 773
\bibitem{kraiselburd} Kraiselburd, L., \& Vucetich, H.\ 2012, Physics Letters B, 718, 21 
\bibitem{kraiselburd2} Kraiselburd, L., Landau, S.~J., Negrelli, C., \& Garcia-Berro, E.\ 2015, Astrophys.\ Space Sci.\  {  357}, no. 1, 4 (2015)
\bibitem{king} King, J.~A., Webb, J.~K., Murphy, M.~T., et al.\ 2012, \mnras, 422, 3370 
\bibitem{mantz}Mantz A., Allen S. W., Ebeling H., Rapetti D., MNRAS, 2008, 387, 1179
\bibitem{mel}Melchiorri, F. \& Melchiorri, B. O., 2005, Proceedings of the International School of Physics “Enrico Fermi”, 159, 225
\bibitem{wei} Jun-Jie, W.,  Xue-Feng, Wu \&   Fulvio, M., 2015, MNRAS, 447, 479
\bibitem{Lima:2003dd} Lima, J.~A.~S., Cunha, J.~V. and Alcaniz, J.~S., 2003, Phys.\ Rev.\ D {68}, 023510
\bibitem{lubini}Lubini, M., Sereno, M., Coles, J., Jetzer, Ph., Saha, P., MNRAS, 2013, 437, 2461
\bibitem{mar}  Martins, C. J. A. P.,   Vielzeuf, P. E.,  Martinelli,  M.,  Calabrese, E. \& Pandolfi, S., 2015, PLB, 743, 377
\bibitem{mol} Molaro, P., {\it{et al.}}, 2013,  A\&A, 555,  A68
\bibitem{mor}Moresco, M., Cimatti, A., Jimenez, R. {\it{et al.}} 2012, JCAP,08, 006
\bibitem{mosquera} Mosquera, M.~E., \& Civitarese, O.\ 2013, \aap, 551, A122 
\bibitem{mu} Muller, J., Hofmann, F. \& Biskupek, L., 2012, CQG, 29, 184006 
\bibitem{murphy} Murphy, M.~T., Webb, J.~K., \& Flambaum, V.~V.\ 2003, \mnras, 345, 609 
\bibitem{no}Nozawa, S., Itoh, N., \& Kohyama, Y. 1998, ApJ, 508, 17-24
\bibitem{not} Noterdaeme, P., {\it{et al.}}, 2011 , A\&AL, 526, L7
\bibitem{obryan} O'Bryan, J., Smidt, J., De Bernardis, F., \& Cooray, A.\ 2015, \apj, 798, 18 
\bibitem{pen}Pen U., New Astronomy , 1997,  2, 309
\bibitem{peik} Peik, E., Lipphardt, B., Schnatz, H., et al.\ 2004, Physical Review Letters, 93, 170801 
\bibitem{petrov} Petrov, Y.~V., Nazarov, A.~I., Onegin, M.~S., Petrov, V.~Y., \& Sakhnovsky, E.~G.\ 2006, \prc, 74, 064610 
\bibitem{planckcol} Planck Collaboration, Ade, P.~A.~R., Aghanim, N., et al.\ 2015, \aap, 580, A22 
\bibitem{re} Renzini, A., 2006,  A.ARA\&A,  44, 141
\bibitem{r}Riess, A.G., {\it{et al.}} 2011, ApJ, 730, 119
\bibitem{ros} Rosenband, T., 2008, Science, 319, 1808
\bibitem{sar}Sarazin, C. L. 1988 in “X-ray emission from clusters of galaxies” Cambridge Astrophysics Series, Cambridge University Press
\bibitem{saro} Saro, A., {\it{et al.}} 2014, MNRAS,  440, 2610
\bibitem{sa}Sasaki S., PASJ, 1996, 48, L119
\bibitem{son}Songaila, A. \&  Cowie,L., 2014, Astrophys.J. 793, 103
\bibitem{sun}Sunyaev, R. A., \& Zel’dovich, Ya.B. 1972, Comments Astrophys. Space Phys., 4, 173
\bibitem{su}Suzuki, N. et al., 2012, ApJ, 85, 746
\bibitem{van}Vanderlinde K., Crawford T. M., de Haan T., Dudley J. P., {\it{et al.}} {\it{et al.}} L. S., ApJ, 2010, 722,  1180
\bibitem{uzan2011}Uzan, J.-P.\ 2011, Living Reviews in Relativity, 14,  
\bibitem{uzan}Uzan, J. P., Aghanim, N., \& Mellier, Y. 2004, Phys. Rev. D, 70, 083533
[astro-ph/0405620v1]
\bibitem{tor} Wagner, T. A.,  Schlamminger, S.,  Gundlach, J. H. \& Adelberger, E. G., 2012, CQG, 29, 184002 
\bibitem{webb} Webb, J.~K., Flambaum, V.~V., Churchill, C.~W., Drinkwater, M.~J., \& Barrow, J.~D.\ 1999, Physical Review Letters, 82, 884 
\bibitem{whitmore} Whitmore, J.~B., \& Murphy, M.~T.\ 2015, \mnras, 447, 446 




\end{thebibliography}
\end{document}